%% file: EffectsOfBottleneck.tex
\documentclass{jpsj2}
\title{Effects of Bottlenecks on Vehicle Traffic}
\author{Syohei Yamamoto$^{1}$, Yasuhiro Hieida$^{2}$ and Shin-ichi Tadaki$^{2}$ }
\inst{$^{1}$Department of Information Science, Saga University, Saga 840-8502, Japan\\
$^{2}$Computer and Network Center, Saga University, Saga 840-8502, Japan}
\abst{Traffic congestion is usually observed at the upper stream of bottlenecks such as tunnels.  
Congestion appears as stop-and-go waves and high-density uniform flow.  We perform simulations of traffic flow with a bottleneck using the coupled map optimal velocity model.  The bottleneck is expressed as a road segment with speed reduction.  
The speed reduction in the bottleneck controls the emergence of stop-and-go waves.
A phenomenological theory of bottleneck effects is constructed. }
\kword{traffic flow, bottleneck, optimal velocity, coupled map}\renewcommand{\refname}{}
\begin{document}
\maketitle
\section{Introduction} 
Traffic flow phenomena have been attracting scientific and engineering research interests since the popularization of cars in the 1950s.  
The physical understanding of traffic flow on expressways has been improved mainly on the basis of mathematical models and their computer simulations since the early 1990s.\cite{1,2}
Many interesting features have been studied from the viewpoints of nonequilibrium statistical physics, pattern formation and transportation phenomena.

One interesting feature observed in traffic flow is the emergence of traffic congestion. 
Near a bottleneck, we observe high-density flow, which shows complex behavior in a density-flux diagram.\cite{3}
The high-density flow breaks down to stop-and-go waves at a distance from the bottleneck. It is pointed out, however, that a bottleneck is not the origin of congestion.\cite{4,5}
A bottleneck just increases the density of traffic flow. 
If the induced density is low, cars run smoothly.
The uniform flow beyond a critical density is unstable and breaks down to stop-and-go waves.
Sugiyama and Nakayama reproduced this feature using computer simulations.\cite{6} 
Mitarai and Nakanishi discussed that convective instability is closely related to the break-down to stop-and-go waves.\cite{7,m2,m3}
The increase in density due to the effect of a bottleneck is the key to understanding the emergence of congested flow at the upper stream of a bottleneck. 
In this study, we perform simulations of a system with a bottleneck under open boundaries and discuss the effect of the bottleneck on the emergence of congestion.

Physical models of traffic flow are, in general, divided into two types; macroscopic and microscopic. 
The macroscopic models treat traffic flow as fluid.  
The microscopic models treat individual cars as particles and describe interactions among them.  
One of the microscopic models is the Nagel-Schreckenberg model\cite{8}, which
is a cellular automaton model of traffic flow.  Another type of microscopic model is the car-following model.\cite{1}

The optimal velocity (OV) model\cite{9} of traffic flow is one of the car-following models.  
The most important feature of the model is the introduction of the optimal velocity.  In the OV model, each car controls its speed to fit the optimal velocity, which is decided by the headway distance to its preceding car.  
The model is described as a set of differential equations for the positions of cars.  
The model is suitable for treating the instability of the high-density traffic flow at the upper stream of a bottleneck such as a tunnel.\cite{4}

We construct a simulation system for observing the emergence of congestion near a bottleneck.  
The system should be an open-road system with the injection and ejection of cars.  So, we employ the coupled map optimal velocity (CMOV) model of traffic flow\cite{10}, which is a temporal discretization of the OV model.  The CMOV model is suitable for computer simulations with open boundaries.

The organization of this paper is as follows:  First, we describe the CMOV model and the setup of the simulations in \S2.  
We construct a one-lane open-road system with a bottleneck.  The bottleneck is implemented as a road segment with speed reduction and suppresses the flux in the bottleneck indirectly. 
The simulation results are shown in \S3. Typically, stop-and-go waves are observed at a distant upper stream of the bottleneck.  
Near the bottleneck there is uniform traffic flow.  
We summarize the relation of the speed reduction in the bottleneck to the appearance of the stop-and-go waves.
A phenomenological theory of bottleneck effects is discussed in \S4.  Section 5 is devoted to summary and discussion.

\section{Model and Simulation Setup}

We employ the CMOV traffic flow model\cite{10},
which is a temporal discretization of the OV model\cite{9}. 
The CMOV model updates the position $x(t)$ and the speed $v(t)$ of a car by
\begin{align}
&x(t+\Delta t)=x(t)+v(t)\Delta t \label{cmovx},\\
&v\left(t+\Delta t\right)=
v(t)+\alpha\left(V_\mathrm{optimal}\left(\Delta x\right)-v(t)\right)
\Delta t,\label{cmovv}
\end{align} 
where $\Delta x$ is the headway distance to the preceding car, $\mathrm{\Delta}t$ is a discrete-time unit given as 0.1~s in this paper,
and $\alpha$ is a sensitivity constant. 
Each car controls its speed to fit 
the optimal velocity decided by the OV function $V_\mathrm{optimal}(\mathrm{\Delta} x)$, which depends on the headway distance $\mathrm{\Delta}x$ to the preceding car.
The OV function is, in general, a sigmoidal function of the headway distance. For realistic simulations, we use the following form:
\begin{align}
V_\mathrm{optimal}\left(\Delta x\right)
=\frac{v_\mathrm{max}}{2}
\left[\tanh\left(2\frac{\Delta x-d}{w}\right)+c\right],\label{ovf}
\end{align}
where parameters $v_\mathrm{max}, d, w$ and $c$ can be obtained through observations of the car-following behavior.
We use the set of the parameters in Table \ref{parameter}, which is compatible with that in ref. \citen{11}.

\begin{table}[h]
		\begin{center}
			\begin{tabular}[htb]{c|c|c} \hline
parameter&value&unit\\ \hline
$d$&25.0&m \\
$w$&23.3&m \\
$v_{\mathrm{max}}$&33.6&m/s\\
$\alpha$&$2.0$&$\mathrm{s}^{-1}$\\
$c$&0.913&\\\hline
			\end{tabular} 
	\caption{Parameters in optimal velocity function eq. (\ref{ovf}).}\label{parameter}
		\end{center}
\end{table}

Cars should stop to avoid backward motion and collision with preceding cars.
The optimal velocity is negative if the headway $\Delta x$ is less than $\Delta x_\mathrm{min}$ which satisfies $V_\mathrm{optimal}(\Delta x_\mathrm{min})=0$.
The avoidance is expressed by replacing eqs. (\ref{cmovx}) and (\ref{cmovv}) with 
\begin{align}
 &x(t+\mathrm{\Delta}t)=x(t),\label{cmovx2}\\
 &v(t+\mathrm{\Delta}t)=0,\label{cmovv2}
 \end{align}
 for $\mathrm{\Delta} x < \mathrm{\Delta} x_\mathrm{min}$.
 
\begin{figure}\begin{center}
\input{system}
\caption{Schematic view of system. Cars are injected from the left side and ejected away from the right side. A bottleneck region is located at the right end of the system. In our simulations, the length of the system and that of the bottleneck region are $L = 10000$(m) and $L_\mathrm{B} = 2000$(m), respectively. }

\label{system}
\end{center}
\end{figure}
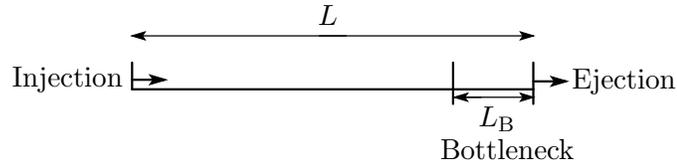

We construct a one-lane road of length $L$ with open boundaries (Fig. \ref{system}). 
 If a car arrives at the right end of the system, it is ejected from the system.  
The headway of a car following the car ejected from the system is set to be $L$ as a headway sufficiently long.
At the left end of the system, a car with zero velocity is injected if the distance between the left end of the system and the tail of the sequence of cars is larger than $\mathrm{\Delta} x_\mathrm{min}$.

We also introduce a bottleneck region of length $L_\mathrm{B}$ at the right end of the system, for observing its effect.  The bottleneck is defined by reducing the maximum speed in the region.  
Namely, cars in the bottleneck run with the reduced OV function $V^{\mathrm{(b)}}_\mathrm{optimal}$:

\begin{align}
V^{\mathrm{(b})}_\mathrm{optimal}\left(\Delta x\right)
=rV_\mathrm{optimal}\left(\Delta x\right),\label{ovfb}
\end{align}
where $r$ ($0 \le r \le 1$) is the degree of speed reduction in the bottleneck.

\section{Simulation Results}

\begin{figure}[htb]\begin{center} 
				\includegraphics[scale=0.4]{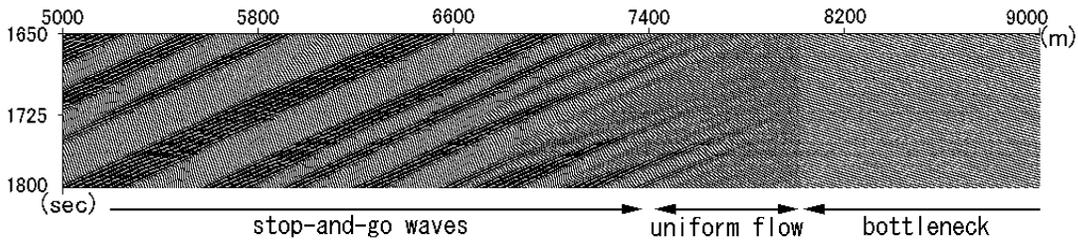}\caption{Space-time plot of car trajectories with intermediate speed reduction $r=0.6$. The horizontal axis denotes 
the positions of cars. The vertical axis denotes the time. The arrows represent regions of stop-and-go waves, uniform flow and the bottleneck, respectively.}\label{timespace}
				\end{center}	
				\end{figure}

We performed simulations with $L=10000(\mathrm{m})$ and $L_\mathrm{B}=2000(\mathrm{m})$.  After relaxation, 
 we can see  typical car trajectories in the space-time plane (Fig. \ref{timespace}).  A high-density uniform region stably exists just before the bottleneck and maintains its length.  
The region is followed by striped patterns which correspond to stop-and-go waves.
They propagate upstream, opposite to the direction of cars. No traffic jam emerges in the bottleneck.
This feature was observed by simulation\cite{6}. Breakdown to stop-and-go waves was discussed as the convective instability of uniform flow without bottlenecks.\cite{7,m2,m3}

\begin{figure}[htb]\begin{center} 
				\includegraphics[scale=0.4]{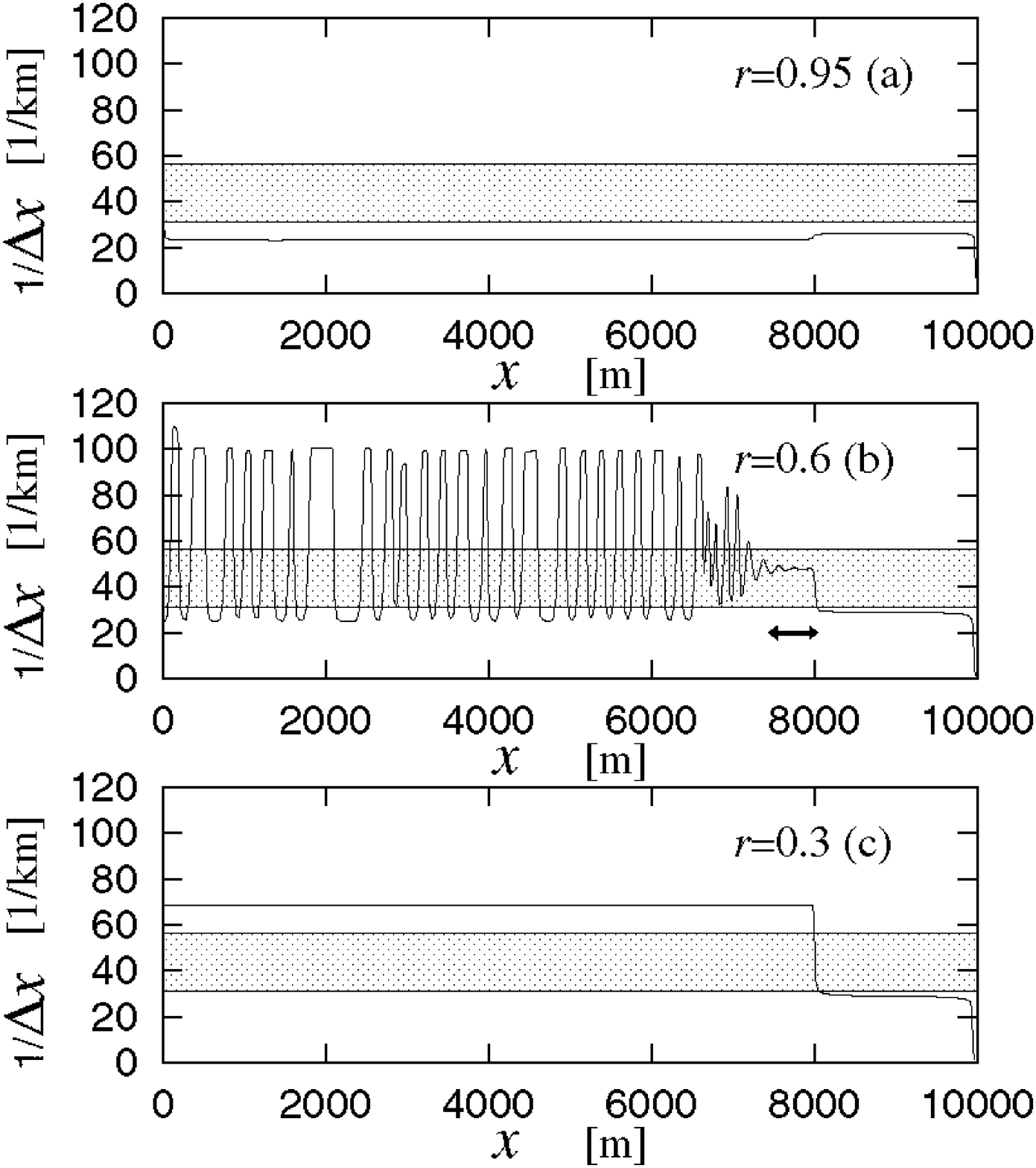}\caption{Snapshots of car density for $r=0.95$ (a), $r=0.6$ (b) and $r=0.3$ (c). 
The horizontal axes denote the positions of cars $x$ (m). Each vertical axis denotes the inverse of headway $1/\Delta x$ (1/km). 
Each of the hatched areas corresponds to the headway with which uniform flow is linearly unstable.
The uniform flow near the bottleneck indicated by an arrow, which corresponds to the uniform flow region in Fig. \ref{timespace}, breaks down to stop-and-go waves, as in case (b). 
}\label{rhox}
				\end{center}	
				\end{figure}
				
The inverse of the headway $1/\Delta x$ is plotted for each car as a snapshot in Fig. \ref{rhox}.
The bottleneck induces three typical patterns of traffic flow.
The emergence of stop-and-go waves depends on headways of cars just before entering the bottleneck.
We can analytically discuss the linear stability of uniform flow using the OV model.\cite{4}
The hatched areas in Fig. \ref{rhox} show the density with which uniform flow is linearly unstable,
2$V'_\mathrm{optimal}(\Delta x)>\alpha$.

\begin{figure}[htb]\begin{center} 
				\includegraphics[scale=0.4]{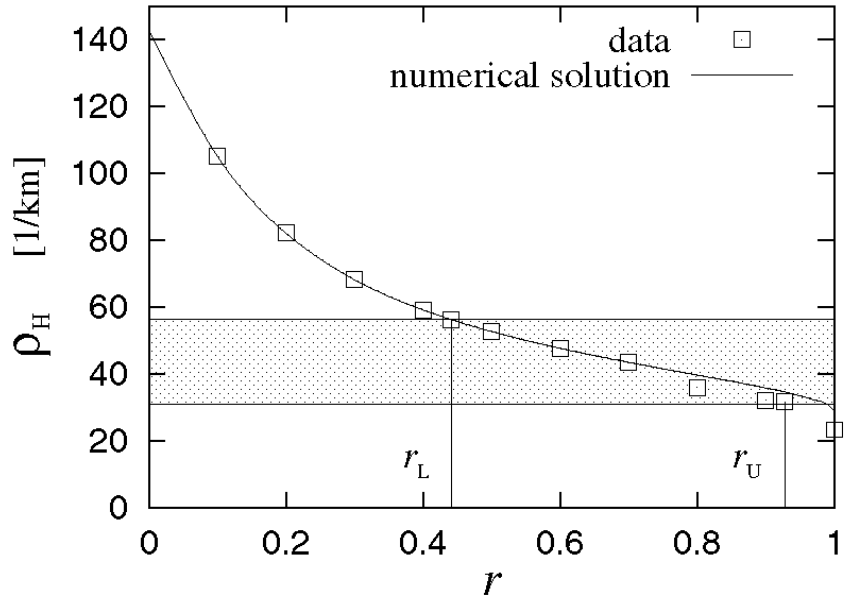}\caption{Relation between speed reduction $r$ and density $\rho_\mathrm{H}$ (1/km) observed at 7800~m point. 
The observed values are shown as $\square$.
The hatched area corresponds to the density at which uniform flow is linearly unstable. The curve is given by our phenomenological theory discussed in \S 4.
The speed reduction between $r_\mathrm{L}\simeq0.44$ and $r_\mathrm{U}\simeq0.92$ induces the density  within the hatched region. }\label{steady}
\end{center}
\end{figure}

After relaxation for 2 h (72000 time steps), we calculate the average density at the 7800~m point by observing the flux and the average of the velocity per hour (36000 time steps) for each $r$ value.
The dependence of the average density $\rho_\mathrm{H}$ on the speed reduction $r$ is shown by the symbols $\square$ in Fig. \ref{steady}. 
We define $r_\mathrm{L}$ and $r_\mathrm{U}$ as the lower and upper bounds, respectively, between which $\rho_\mathrm{H}$ remains in the hatched region.
The boundary values $r_\mathrm{L}$ and $r_\mathrm{U}$ of the speed reduction are obtained as approximately 0.44 and 0.92, respectively, using the simulations. 
These values, however, are slightly different from the boundary values of the emergence of stop-and-go waves in the simulations. The discrepancy comes from the temporal discreteness of the CMOV model and the finiteness of the system length employed in this study.

\begin{figure}[htb]\begin{center} 
\includegraphics[scale=2.0]{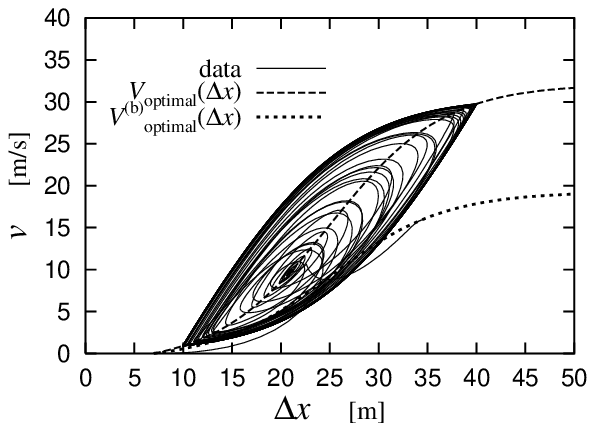}\caption{Typical motion of test car in plane of headway $\Delta x$(m) and velocity $v$(m/s) for case with intermediate speed reduction $r=0.6$ ($r_\mathrm{L}<r<r_\mathrm{U}$).  
The car trajectory draws a hysteresis loop first. The trajectory converges to a point on the curve of the OV function $V_\mathrm{optimal}(\mathrm{\Delta}x)$ as the car approaches the bottleneck.}\label{dxv}
\end{center}\end{figure}

A test car is injected from the left end of the system to observe its behavior. 
Its typical trajectory in the headway-velocity plane is shown in Fig. \ref{dxv} for intermediate speed reduction ($r_\mathrm{L}<r=0.6<r_\mathrm{U}$). 
First, the trajectory draws a closed loop called the \textit{hysteresis loop} while the car continues the stop-and-go motion during the approach to the bottleneck. 
As the car approaches the uniform flow region before the bottleneck, the loop converges to a point on the curve of the OV function. 
Namely, the car moves with the optimal velocity given by the OV function. After the car enters the bottleneck, 
the trajectory moves to a point on the curve of the reduced OV function in the bottleneck.
As the car approaches the right end of the system (the end of the bottleneck), the uniformity of the headway in the bottleneck is lost (this part of the trajectory is not shown in Fig. \ref{dxv}).

\begin{figure}[htb]\begin{center} 
				\includegraphics[scale=2.0]{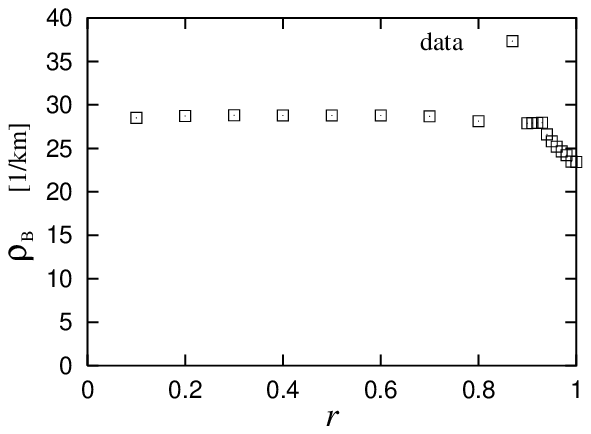}\caption{Relation between density $\rho_\mathrm{B}$ (1/km) in bottleneck (at 9000m point) and speed reduction $r$. 
Except in very weak speed reduction cases $r>r_\mathrm{U}\simeq 0.92$, the density is independent of speed reduction.}\label{steady2}
\end{center}
\end{figure}

\begin{figure}[htb]\begin{center} 
				\includegraphics[scale=2.0]{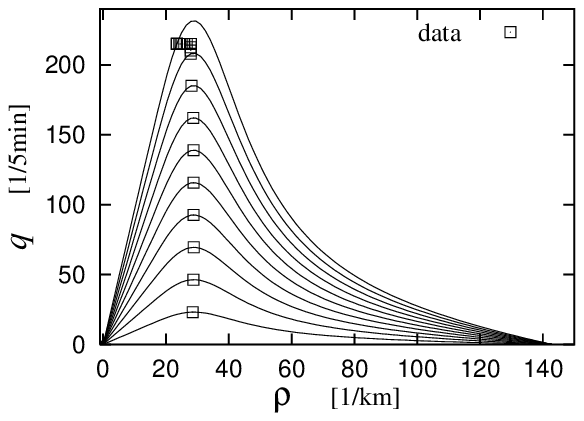}\caption{Curves denoting relation eq. (\ref{homo}) between flux $q$ (1/5min) and density $\rho$ (1/km) of uniform flow in bottleneck (at 9000m point) for various values of speed reduction $r$.
The curves correspond to $r=1.0, 0.9, 0.8,\cdots, 0.2, 0.1$ (from top to bottom), respectively.
The observed values of $\rho_\mathrm{B}$ of the flow are shown as $\square$. These values of $\rho_\mathrm{B}$ correspond to those in Fig. \ref{steady2}.}\label{steady3}
\end{center}
\end{figure}

The relation between $r$ and the density in the bottleneck, $\rho_\mathrm{B}$, is shown in Fig. \ref{steady2}.
$\rho_\mathrm{B}$ is observed at the 9000m point by the same method as $\rho_\mathrm{H}$. 
Except for the very weak speed reduction ($r > r_\mathrm{U}$), $\rho_\mathrm{B}$ is independent of $r$.
We interpret these observed results in the density-flux relation. A uniform flow is observed in the bottleneck. 
We can calculate the flux $q$ of the uniform flow with the optimal velocity in the bottleneck as a function of the density $\rho$:
\begin{align}
 q&=\rho V_\mathrm{optimal}^{(\mathrm{b})} \left(\frac{1}{\rho}\right).\label{homo}
 \end{align}
Figure \ref{steady3} shows the relation between $q$ obtained from eq. (\ref{homo}) and the observed $\rho_\mathrm{B}$.
$\rho_\mathrm{B}$ corresponds to the maximum flux in the bottleneck.

\section{Phenomenological Theory of Bottleneck Effect}

\begin{figure}\begin{center}
\input{rho_HB}
\caption{Schematic diagram of effect of bottleneck.}
\label{entering}
\end{center}
\end{figure}
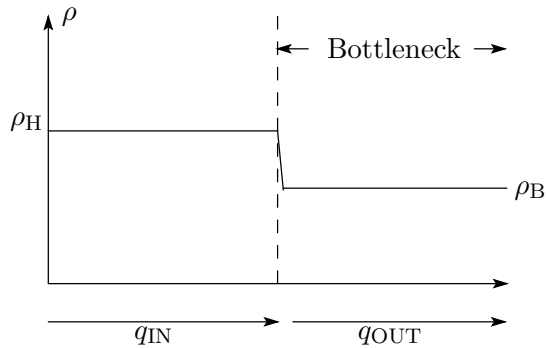

We are interested in the effect of a bottleneck on the traffic flow at the upper stream of a bottleneck. 
Here, we construct a phenomenological theory of the bottleneck effect. We make two assumptions on the basis of our simulations. 
The first one is that a uniform-density flow exists just before the bottleneck.
Thus, cars move with the optimal velocity just before entering the bottleneck. 
The flux $q_\mathrm{IN}$ entering the bottleneck is given by a function of $\rho_\mathrm{H}$:

\begin{align}
q_\mathrm{IN}(\rho_\mathrm{H})=\rho_\mathrm{H}V_\mathrm{optimal}\left(\frac{1}{\rho_\mathrm{H}}\right),
\end{align}
where $\rho_\mathrm{H}$ is the density of cars just before the bottleneck (Fig. \ref{entering}). 

Second, after entering the bottleneck, $\rho_\mathrm{B}$ 
is tuned to give the maximum flux: The flux in the bottleneck $q_\mathrm{OUT}$ is given as a function of $r$ by 
\begin{align}
q_\mathrm{OUT}(r)&=\rho_\mathrm{B}V^{\mathrm{(b)}}_\mathrm{optimal}\left(\frac{1}{\rho_\mathrm{B}}\right)=\max_{\rho} \rho V^{\mathrm{(b)}}_\mathrm{optimal}\left(\frac{1}{\rho}\right),\\
\rho_\mathrm{B}&=\underset{\rho}{\mathrm{argmax}}\ \rho V_\mathrm{optimal}^{(\mathrm{b})} \left(\frac{1}{\rho}\right). \label{rhob}
\end{align}

The conservation law of flux requires that the flux $q_\mathrm{IN}$ entering the bottleneck is equal to $q_\mathrm{OUT}$:
\begin{align}
q_\mathrm{IN}(\rho_\mathrm{H})=q_\mathrm{OUT}(r).\label{law}
\end{align}
Using this equation, we obtain $\rho_\mathrm{H}$ just before the bottleneck as a function of $r$.

The effect of the bottleneck is shown as the curve in Fig. \ref{steady} by solving eq. (\ref{law}) numerically. The curve 
describes well the simulation results except for very weak speed reduction $r>r_\mathrm{U}\simeq 0.92$.  From the curve, we obtain two boundary values $r_\mathrm{L}'\simeq 0.441$ and $r_\mathrm{U}'\simeq 0.989$ of the speed reduction.
The intermediate speed reduction, $r_\mathrm{L}'<r<r_\mathrm{U}'$, induces the car density at which uniform flow is linearly unstable (the hatched area in Fig. \ref{steady}). 
Thus, the stop-and-go waves emerge at a distant upper stream of the bottleneck.  
Therefore, by solving eq. (\ref{law}) numerically, we can predict the occurrence of the stop-and-go waves using the value of $r$.

The numerical value $r_\mathrm{L}'$ of the lower bound agrees well with the simulation value $r_\mathrm{L}$.
On the other hand, the upper bound $r_\mathrm{U}'$ disagrees with the simulation value $r_\mathrm{U}$.
The reason is explained simply. The injection method employed in this study cannot supply the 
maximum flux at the upper stream of the bottleneck. 
The maximum flux corresponding to the speed reduction for $r>r_\mathrm{U}$ exceeds the injected flux.
In other words, the flow injected from the left of the system is not sufficient to supply the maximum flux in the bottleneck. As a result, the assumption in the phenomenological theory is not satisfied for $r>r_\mathrm{U}$.

\section{Summary and Discussion}

We studied the effect of a bottleneck using simulations and a phenomenological theory. We employed the coupled map optimal velocity (CMOV) model for simulations. The bottleneck is defined as a road segment with speed reduction.
We obtained the relation between the speed reduction $r$ and the car density $\rho_\mathrm{H}$ before the bottleneck (Fig. \ref{steady}):
The very weak speed reduction, $r>r_\mathrm{U}\simeq0.92$, does not increase the car density $\rho_\mathrm{H}$ to form the stop-and-go waves (Fig. \ref{rhox}(a)). 
The very strong speed reduction, $r<r_\mathrm{L}\simeq0.44$, increases the density $\rho_\mathrm{H}$, which is high enough to stabilize the uniform flow (Fig. \ref{rhox}(c)).
The bottleneck with the intermediate speed reduction, $r_\mathrm{L}<r<r_\mathrm{U}$, induces the high-density uniform flow (represented by the arrow in Fig. \ref{rhox}(b)) just before the bottleneck.
This uniform flow is linearly unstable and breaks down to stop-and-go waves at the distant upper stream of the bottleneck. 

For the intermediate speed reduction, we find two important features. The first is that cars move with the optimal velocity just before the bottleneck. 
The second is that the bottleneck induces the maximal flux within the bottleneck itself. 

We employ these two features as assumptions for the phenomenological theory of the bottleneck effect.  
Using the conservation law of flux, we obtain the density just before the bottleneck as a function of speed reduction (the curve in Fig. \ref{steady}). 
If the density corresponds to that of the linearly unstable uniform flow, the stop-and-go waves emerge at the distant upper stream of the bottleneck. 
Namely, we can predict the occurrence of the stop-and-go waves using the speed reduction.

The effects of bottlenecks have been studied previously.\cite{3,6,7,m2,m3,12}
The breakdown of high-density flow to stop-and-go waves has been observed in simulations.\cite{6,12}
The breakdown effect has been discussed in relation to the convective instability of uniform flow.\cite{7,m2,m3}
The discussion in this paper, however, is based only on the relation between the speed reduction $r$ and the bounds of the linear instability in ref. \citen{4}. The properties of the convective instability may affect the detailed properties induced by a bottleneck, including the stability of high-density uniform flow near the bottleneck. We will discuss these features elsewhere.

\acknowledgements

A part of this work is financially supported by Grants-in-Aid No. 15607014 and No. 18500215 from the Ministry of Education, Culture, Sports, Science and Technology, Japan.

\end{document}

%% file: system
\unitlength 0.1in
\begin{picture}( 31.6000,  7.0100)(  3.0000,-12.0000)
%
\special{pn 13}%
\special{pa 1158 866}%
\special{pa 1158 1006}%
\special{fp}%
\special{pa 3258 1006}%
\special{pa 3258 866}%
\special{fp}%
\special{pa 1158 1006}%
\special{pa 3258 1006}%
\special{fp}%
\special{pa 2838 1006}%
\special{pa 2838 866}%
\special{fp}%
\put(11.1000,-10.2000){\makebox(0,0)[rb]{Injection}}%
\put(34.6000,-10.3000){\makebox(0,0)[lb]{Ejection}}%
\put(27.7000,-13.7000){\makebox(0,0)[lb]{Bottleneck}}%
%
\special{pn 13}%
\special{pa 2838 1006}%
\special{pa 2838 1082}%
\special{fp}%
%
\special{pn 13}%
\special{pa 3258 1006}%
\special{pa 3258 1082}%
\special{fp}%
%
\special{pn 8}%
\special{pa 2844 1048}%
\special{pa 3258 1048}%
\special{fp}%
\special{sh 1}%
\special{pa 3258 1048}%
\special{pa 3190 1028}%
\special{pa 3204 1048}%
\special{pa 3190 1068}%
\special{pa 3258 1048}%
\special{fp}%
%
\special{pn 13}%
\special{pa 1158 956}%
\special{pa 1326 956}%
\special{fp}%
\special{sh 1}%
\special{pa 1326 956}%
\special{pa 1258 936}%
\special{pa 1272 956}%
\special{pa 1258 976}%
\special{pa 1326 956}%
\special{fp}%
%
\special{pn 13}%
\special{pa 3258 964}%
\special{pa 3426 964}%
\special{fp}%
\special{sh 1}%
\special{pa 3426 964}%
\special{pa 3358 944}%
\special{pa 3372 964}%
\special{pa 3358 984}%
\special{pa 3426 964}%
\special{fp}%
%
\special{pn 8}%
\special{pa 2984 1048}%
\special{pa 2844 1048}%
\special{fp}%
\special{sh 1}%
\special{pa 2844 1048}%
\special{pa 2912 1068}%
\special{pa 2898 1048}%
\special{pa 2912 1028}%
\special{pa 2844 1048}%
\special{fp}%
\put(29.7000,-10.9600){\makebox(0,0)[lt]{$L_\mathrm{B}$}}%
%
\special{pn 8}%
\special{pa 2138 726}%
\special{pa 3258 726}%
\special{fp}%
\special{sh 1}%
\special{pa 3258 726}%
\special{pa 3190 706}%
\special{pa 3204 726}%
\special{pa 3190 746}%
\special{pa 3258 726}%
\special{fp}%
\special{pa 2278 726}%
\special{pa 1158 726}%
\special{fp}%
\special{sh 1}%
\special{pa 1158 726}%
\special{pa 1224 746}%
\special{pa 1210 726}%
\special{pa 1224 706}%
\special{pa 1158 726}%
\special{fp}%
\put(21.3000,-6.6900){\makebox(0,0)[lb]{$L$}}%
\end{picture}%

%% file: rho_HB
\unitlength 0.1in
\begin{picture}( 26.4000, 17.2500)(  8.0000,-18.0500)
%
\special{pn 8}%
\special{pa 1000 1600}%
\special{pa 1000 200}%
\special{fp}%
\special{sh 1}%
\special{pa 1000 200}%
\special{pa 980 268}%
\special{pa 1000 254}%
\special{pa 1020 268}%
\special{pa 1000 200}%
\special{fp}%
\special{pa 1000 1600}%
\special{pa 3400 1600}%
\special{fp}%
\special{sh 1}%
\special{pa 3400 1600}%
\special{pa 3334 1580}%
\special{pa 3348 1600}%
\special{pa 3334 1620}%
\special{pa 3400 1600}%
\special{fp}%
\put(10.7000,-2.5000){\makebox(0,0)[lb]{$\rho$}}%
\put(8.0000,-8.0000){\makebox(0,0)[lb]{$\rho_\mathrm{H}$}}%
%
\special{pn 8}%
\special{pa 1000 1800}%
\special{pa 2200 1800}%
\special{fp}%
\special{sh 1}%
\special{pa 2200 1800}%
\special{pa 2134 1780}%
\special{pa 2148 1800}%
\special{pa 2134 1820}%
\special{pa 2200 1800}%
\special{fp}%
\special{pa 2280 1800}%
\special{pa 3390 1800}%
\special{fp}%
\special{sh 1}%
\special{pa 3390 1800}%
\special{pa 3324 1780}%
\special{pa 3338 1800}%
\special{pa 3324 1820}%
\special{pa 3390 1800}%
\special{fp}%
\put(26.2000,-19.2000){\makebox(0,0)[lb]{$q_\mathrm{OUT}$}}%
\put(14.5000,-19.2000){\makebox(0,0)[lb]{$q_\mathrm{IN}$}}%
%
\special{pn 8}%
\special{pa 1000 800}%
\special{pa 2200 800}%
\special{fp}%
\special{pa 2200 800}%
\special{pa 2230 1110}%
\special{fp}%
\special{pa 2230 1100}%
\special{pa 3400 1100}%
\special{fp}%
%
\special{pn 8}%
\special{pa 2200 1600}%
\special{pa 2200 200}%
\special{da 0.070}%
\put(24.6000,-4.2000){\makebox(0,0)[lb]{Bottleneck}}%
\put(34.4000,-11.6000){\makebox(0,0)[lb]{$\rho_\mathrm{B}$}}%
%
\special{pn 8}%
\special{pa 2360 370}%
\special{pa 2200 370}%
\special{fp}%
\special{sh 1}%
\special{pa 2200 370}%
\special{pa 2268 390}%
\special{pa 2254 370}%
\special{pa 2268 350}%
\special{pa 2200 370}%
\special{fp}%
%
\special{pn 8}%
\special{pa 3230 370}%
\special{pa 3390 370}%
\special{fp}%
\special{sh 1}%
\special{pa 3390 370}%
\special{pa 3324 350}%
\special{pa 3338 370}%
\special{pa 3324 390}%
\special{pa 3390 370}%
\special{fp}%
\end{picture}%

%% file: EffectsOfBottleneck.bbl
\begin{thebibliography}{99} 
\bibitem{1}D. Chowdhury, L. Santen and A. Schadschneider: Phys. Rep. \textbf{329} (2000) 199.
\bibitem{2}S. P. Hoogendoorn, S. Luding, P. H. L. Bovy, M. Schreckenberg and D. E. Wolf, ed.: \textit{Traffic and Granular Flow '03} (Springer, Berlin, 2005) p. 305.
\bibitem{3}B. S. Kerner and H. Rehborn: Phys. Rev. E \textbf{53} (1996) R4275.
\bibitem{4}M. Bando, K. Hasebe, A. Nakayama, A. Shibata and Y. Sugiyama: 
Phys. Rev. E \textbf{51} (1995) 1035.
\bibitem{5}Y. Sugiyama, A. Nakayama, M. Fukui, K. Hasebe, M. Kikuchi, K. Nishinari, S. Tadaki and S. Yukawa: \textit{Traffic and Granular Flow '03} (Springer, Berlin, 2005)p. 45.
\bibitem{6}Y. Sugiyama and A. Nakayama: AIP Conf. Proc. \textbf{661} (2003) 111. 
\bibitem{7}N. Mitarai and H. Nakanishi: J. Phys. Soc. Jpn. \textbf{68} (1999) 2475. 
\bibitem{m2}N. Mitarai and H. Nakanishi: Phys. Rev. Lett. \textbf{85} (2000) 1766.
\bibitem{m3}N. Mitarai and H. Nakanishi: J. Phys. Soc. Jpn. \textbf{69} (2000) 3752.  
\bibitem{8}K. Nagel and M. Schreckenberg: J. Phys. I (France) \textbf{2} (1992) 2221.
\bibitem{9}M. Bando, A. Nakayama, A. Shibata and Y. Sugiyama: Jpn. J. Ind. Appl. Math. \textbf{11} (1994) 203.
\bibitem{10}S. Tadaki, M. Kikuchi, Y. Sugiyama and S. Yukawa: J. Phys. Soc. Jpn. \textbf{67} (1998) 2270. 

\bibitem{11} M. Bando, K. Hasebe, A. Nakayama, A. Shibata and Y. Sugiyama: J. Phys. I (France) \textbf{5} (1995) 1380.
\bibitem{12} M. Kikuchi, Y. Sugiyama, S. Tadaki and S. Yukawa: \textit{Control In Transportation Systems 2003} (Elsevier, Oxford, 2004) p. 347.
\end{thebibliography}
